\documentclass[aps,prl,a4paper,preprintnumbers,nofootinbib,twocolumn]{revtex4}
\pdfoutput=1
\usepackage{amsmath,bbm,latexsym,url,graphicx}
\usepackage[utf8]{inputenc}

\renewcommand{\Im}{{\rm Im}\,}
\renewcommand{\Re}{{\rm Re}\,}
\newcommand{\re}[1]{\Re\left(#1\right)}
\newcommand{\im}[1]{\Im\left(#1\right)}
\newcommand{\mathds}[1]{}

\renewcommand{\sp}[2]{\left\langle  \vphantom{#1}\vphantom{#2} #1\right.\left.\vphantom{#1}\vphantom{#2} \right| \left. #2  \vphantom{#1}\vphantom{#2} \right\rangle}

\newcommand{\K}{\mathcal{K}}
\newcommand{\R}{\mbox{$\mathcal{R}$}}
\newcommand{\I}{\mbox{$\mathcal{I}$}}
\newcommand{\Iinf}{\mbox{$\mathcal{I_\infty}$}}

\newcommand{\Pp}{\text{P}_+}
\newcommand{\Pm}{\text{P}_-}
\newcommand{\Ppm}{\text{P}_\pm}

\newcommand{\vecr}{\mathbf{x}}

\newcommand{\ssigma}{\mbox{${\cal S}$}}
\newcommand{\ssigmai}{\ssigma}
\newcommand{\atot}{\mbox{$A$}}

\newcommand{\ibar}{{\bar \imath}}
\newcommand{\jbar}{{\bar \jmath} }
\newcommand{\abar}{{\bar a}}

\newcommand{\kbar}{{\bar k}}

\begin{document}
\title{Smarr Mass formulas for BPS multicenter Black Holes }

%\author{ J.J.~Fernandez-Melgarejo}\email[]{ jj.fernandezmelgarejo@um.es}
\author{ E.~Torrente-Lujan}\email[]{ torrente@cern.ch}

\affiliation{ Dept. de F\'{\i}sica, U. de Murcia, Campus de Espinardo, 30100 Murcia, Spain}

\affiliation{TH division, CERN, 1211-Geneve-23, CH}

%\date{}
\preprint{FISPAC-TH/19-314}
\preprint{UQBAR-TH/19-156}

\begin{abstract}
Mass formulas for multicenter BPS 4D black holes are presented. For example, ADM mass for a two center BPS solution can be related to
the intercencenter distance $r$, the angular momentum $J^2$, the dyonic charge vectors $q_i$ and 
the value of the scalar moduli at infinity ($z_\infty$)by the relation
$M_{ADM}^2 =A\left (1+ \alpha J^2\left(1+\frac{2M_{ADM}}{r}+\frac{A}{r^2}\right)\right)$
where $A(Q),\alpha(q_i)$
are symplectic invariant quantities
($Q$, the total charge vector) 
depending on the special geometry prepotential defining the theory.
The formula predicts the existence of a continuos class, 
for fixed value of the charges, 
of BH's with interdistances $r\in (0,\infty)$ and   $M_{ADM}\in (\infty,M_\infty)$.
 Smarr-like expressions incorporating the 
intercenter distance are obtained from it:
$$ dM\equiv\Omega d J+\Phi_i d q_i+ F dr,$$
in addition to an effective angular velocity $\Omega$ and 
electromagnetic potentials $\Phi_i$, the equation 
allows to define an effective ``force'', $F$,
 acting between the centers. This  effective force is always negative:
at infinity we recover the familiar Newton law $F\sim 1/r^2$
while at short distances $F\sim f_0+f_1/r^2$. Similar results can be easily obtained for more general models and number of 
centers.
\end{abstract}

\maketitle

\noindent
{\bf 1. Introduction.} 
Extended theories of gravity as Supergravity, 
and in particular supersymmetric 
extremal black hole solutions,  continues to be 
central for M-theory, string theory phenomenology, 
quantum properties of black holes, 
and the AdS/CFT correspondence. 
Applications can be found  from extensions of the 
particle physics SM to supersymmetric black hole solutions 
and strongly coupled systems. %\cite{}.
Black hole physics, 
in general relativity or in extended theories as supergravity, 
is of interest in different 
backgrounds:
from astrophysics to classical general relativity,
quantum field theory, particle physics, string 
and supergravity.
Nowadays black holes in Supergravity theories 
are used  to answer 
to condensed matter questions in particular in 
strongly correlated fermionic 
systems including high T superconductors
\cite{Lee:2008xf,Cubrovic:2010bf,Hartnoll:2007ih,Macher:2009nz,Lahav:2009wx,Gentle:2013pua,Dvali:2012rt}.

It is a rather trivial problem the existence and 
construction of extremal BH solutions in a wide number 
of well known theories.
The two-parameters  Reissner-Nordstrom (RN) metric,
for example, describes black holes 
of (ADM) mass $M$ and charge $Q$ only when the ratio $Q/M$ 
is sufficiently small. 
%For large $Q$ it describes a naked singularity. 
In the extremal case, the borderline between
naked singularities and black hole solutions, the mass and
and electromagnetic charges $P,Q$ are related by 
$M^2= P^2+Q^2.$
This can be considered a (neccesary and suficient) condition
on the macroscopic parameters for the existence of a 
 extremal RN BH.
Solutions saturating this bound can be considered as the 
stable final state of Hawking evaporation
\cite{pioline:clasquangrav23-2006}.
%
%The horizon area is given in this extremal case by $A_{h,extr}=4\pi M^2$.
%
For the Majumdar-Papatreou solution with
$H=1+\sum_i M_i/\lvert x-x_i\rvert$, the conditions are $M_i>0$ with
$M_i^2=q_i^2$,  $M_{ADM}=\sum M_i$.
For the three parameter Kerr-Newman extremal case, 
the mass, charge and
angular momentum of the solution are related by the 
(quartic in the mass) extremal condition
$M^2=Q^2+J^2/M^2.$ %% quartic implies non supersymmetric??.
From this relation we get the limits 
$M^2\simeq Q^2$ for $J\to 0$ and 
$M^2\simeq \lvert J\rvert $ for $J\to \infty$. 
%The horizon area for the extremal case is $A_{h,extr}= 4\pi \left ( 2 M^2-Q^2\right )$ $=4\pi\sqrt{Q^4+4 J^2}$.
%For $J\to 0$, $A_{h,extr}\simeq 4\pi M^2\simeq 4\pi Q^2$.
%
In the supergravity dilaton model, a 4d low energy limit of 
string theory, the presence of the dilaton induces that the
transition between black hole and naked singularities occurs
at \cite{Garfinkle:1990qj}
%% \bibitem{Garfinkle:1990qj}D.~Garfinkle, G.~T.~Horowitz and A.~Strominger,  %``Charged black holes in string theory,''Phys.\ Rev.\ D {\bf 43} (1991) 3140   [Erratum-ibid.\ D {\bf 45} (1992) 3888].  %%CITATION = PHRVA,D43,3140;%%%757 citations counted in INSPIRE 
$M^2= \frac{1}{2} Q^2 \exp 2\phi_\infty$
where $\phi_\infty$ is the value of the additional scalar 
field at spatial infinity. 
The factor modifying this last formula from the 
RN relation is related to the existence of a dilaton 
scalar charge \cite{Garfinkle:1990qj}.
For the RN metric the extremal value corresponds to the
case where gravitational and electromagnetic interactions 
are balanced. In any supergravity scenario,
nearly unavoidably, the scalars contribute with an extra 
attractive long range force.
In the Maxwell-Einstein axion-dilaton supergravity
\cite{Ferrara:2012qp,Kallosh:1993yg}
extremal one center case
one obtains a \emph{quartic} expression for the ADM mass
%the following quartic expression for the ADM mass
%%\begin{eqnarray}M^4-\Delta_+(Q) M^2+\Delta_-(Q)/4 &=&0\end{eqnarray}with $\Delta_\pm q^2e^{2\phi_\infty}\pm p^2 e^{-2\phi_\infty}$
in terms of $Q,P$, electric and magnetic charges,
and $\phi_\infty$, the value of the dilaton at spatial 
infity (the axion is put to zero in addition).
In the extremal BPS case the relation reduces to
\emph{quadratic} expression
$M^2=\frac{1}{2}\left(p e^{-\phi_\infty}+q e^{\phi_\infty}\right )^2$. 
%
%% XXX ... other mass relations from ortin??
Finally, mass relations  are known for one center 
extremal black hole solutions:
 For $4d, N=2$ ungauged 
SUGRA theories  the following constraint between the 
BH mass, the scalar charges, the scalar metric $G_{ab}$ evaluated at spatial infinity and the BH potential is well known
\cite{Ferrara:1997tw,libroortin}
%%\bibitem{libroortin} T. Ortin, Gravity and Strings, Cambridge 2003.
%
%%\bibitem{Ferrara:1997tw}S.~Ferrara, G.~W.~Gibbons and R.~Kallosh,  %``Black holes and critical points in moduli space,'' Nucl.\ Phys.\ B {\bf 500} (1997) 75  [hep-th/9702103].
  %%CITATION = HEP-TH/9702103;%% %369 citations counted in INSPIRE 
% 9702103 and refs 3,4 therein
$M^2+ G_{ab}^\infty \Sigma^a\Sigma^b-V_{bh}(p,q,\phi^a_\infty)=0$.
%The effective BH potential $V_{bh}$ is given by...
The quantities $\Sigma^a,\phi_\infty^a$ are defined by the 
expansion of the scalar fields at infinity 
$\phi^a(r\to\infty)\sim \phi_\infty^a+\Sigma^a/r+o(1/r^2)$.
%Usually  $\Sigma^a,\phi_\infty^a$ are not independent of each other.
%
In summary,
given a set of parameters 
$M,Q,\phi_\infty^a,J,\Sigma^a..$ (or a subset of them) 
satisfiying simple relations as the previous ones, 
we can build explicitly the solutions from them. 
By  construction such 
relations connecting ADM masses, charges and
possibly other moduli can be considered local neccesary 
and sufficient conditions for the existence of one center 
extremal BH solutions.

%multicenter solutions,existence criteria denef0702146
%
Multicenter BH solutions are an important ingredient on, for example, counting the right numbers of degree of freedom in Entropy BH computations. 
However it is not a trivial problem in general 
the proof of the existence or not 
of BPS multicenter solutions of given charges, center positions and values
of the moduli (scalar fields at spatial infinity).
The most general stationary (time independent) 4-dimensional metric compatible with supersymmetry can be
written in the IWP form
\cite{Israel:1972vx,Perjes:1971gv,Sabra:1997yd},
\begin{align}
 ds^2&=e^{2 U}(dt+\omega)^2-e^{-2 U} d{\vecr}^2.
\label{eq211}
\end{align}
This is in particular the
 metric of a $4d$ BPS solution of general 
$N=2$ supergravity theories coupled to vectors and scalars. 
In terms of this metric, 
some neccesary and sufficient, non local conditions 
for the existence of multicenter extremal solutions are 
\cite{Denef:2007vg}:
%%\bibitem{Denef:2007vg}F.~Denef and G.~W.~Moore,
  %``Split states, entropy enigmas, holes and halos,''JHEP {\bf 1111} (2011) 129  [hep-th/0702146 [HEP-TH]].
  %%CITATION = HEP-TH/0702146;%%
  %255 citations counted in INSPIRE 
1) the fullfillment of certain integrability conditions for 
the 1-form $\omega$ which imposes 
restrictions on the allowed center positions;
2) the metric factor $e^{2 U}$ is positive  at any 
spacetime point;
3) the scalar field solutios of the theory, $z^a(\vecr)$,
%related to the Kahler potential and the scalar metric 
 must adopt physically consistent values at any spacetime point.
In particular the conditions that the charge vectors at 
any center, $q_i$, define by themselves a single black hole
solution (in particular positivity of the associated entropy)
are neccesary but not sufficient for the existence of a 
multicenter BPS solution corresponding to that set of vectors.

It would be desiderable to have a fully local set of 
neccesary and sufficient BPS existence conditions 
in terms of ``macroscopical'' parameters
$M_{ADM},q_i,\phi_\infty^a$, 
physical parameters appearing, or determining, directly 
in the field equations and their solutions
but this is not known.

%Given a set of parameters $M,Q,\phi_\infty^a,J,\Sigma^a..$ (or a subset of them) satisfiying, in their respective contexts,simple relations as \ref{eqx1} or \ref{eqx2}, we can build explicitly the solutions from them. By  construction, relations as \ref{eqx1} or \ref{eqx2}, can be considered local neccesary and sufficient conditions for the existence of one center extremal BH solutions.

%...It would be convenient to have neccesary and sufficient conditions expressed in terms of the macroscopic parameters $M,Q,\phi_\infty^a,J,r_{ij},\Sigma^a..$.

Supersymmetric $N=2$ supergravity solutions, in particular
multicenter stationary BPS solutions, can be constructed
systematically following well established methods
\cite{Bates:2003vx,Meessen:2006tu,ferraragimonkallosh74prd2006,Bellorin:2006xr,Fernandez-Melgarejo:2013ksa}\footnote{We refer to \protect\cite{Bellorin:2006xr,Fernandez-Melgarejo:2013ksa} for further details and notation fixing.}
The 1-form $\omega$ and the function $e^{-2U}$ are related in
these theories to the K\"ahler potential and
connection, $\K,Q$ \cite{Sabra:1997yd}.
%%K\"ahler gauge fixing is accomplished by
We demand asymptotic flatness, $e^{-2 U}\to 1$ together with $\omega\to 0$ for $|\vecr|\to\infty$.
BPS field equation solutions  can be written
in terms of real symplectic vectors $\R$ and $\I$.
For example
$e^{-2 U} =  \sp{\R}{\I}$. %=\sp{\ssigma\I}{\I}\, .
%\label{eq215}\end{align}
%%\begin{align}
%%\R&=\frac{1}{\sqrt 2}\re{\frac{V}{ X}} \, ,\\
%%\I&=\frac{1}{\sqrt 2}\im{ \frac{V}{ X}}\, .
%%\end{align}$X$ is an arbitrary complex function of space coordinates such that $1/X$ is harmonic.
The $2 n_v+2$ components of $\I$ and $\R$ are 
${R}^3$ harmonic functions. 
There is an algebraic  relation between $\R$ and $\I$, the 
\emph{stabilization equation}
\begin{eqnarray}
\R &=& \ssigma\I\, ,
\label{eq888}
\end{eqnarray}
where the $\ssigma$ matrix is given in 
terms of the second derivatives of the, assumed quadratic, prepotential defining
the theory\footnote{The matrix $S(z)$ is nothing else 
the matrix ${\cal M}(F)$ appearing in \protect\cite{Ceresole:1995ca,Fernandez-Melgarejo:2013ksa} 
The bilinear form $\sp{S q_i}{q_j}$ is not the usually defined 
$I_1$ invariant, but coincides with it at the attractors points
(infinity and charge centers). See \protect\cite{Fernandez-Melgarejo:2013ksa}.}. .

Similarly,  the time independent $3$-dimensional 1-form  $\omega=\omega_i dx^i $ satisfies the equation
 $d\omega=2\sp{\I}{\star_3 d\I}$
% \label{eq687bb}\end{align}
together with  the integrability condition
$ \sp{\I}{\Delta\I}=0.$
%\label{eq344}\label{eq221}\end{align}
As a consequence of this condition, 
the center interdistances are restricted, 
($r_{ab}=| \vecr_a-\vecr_b |$, for any $q_b$ )
\begin{eqnarray}
\sp{\I_\infty}{q_b}
+\sum_a\frac{\sp{q_a}{q_b}}{r_{ab}} &=&0.
\label{eq5554}
\end{eqnarray}
The previous 
equations implies $N=\sum N_b=0$
where  $N_b=\sp{\I_\infty}{q_b}$ are the `` NUT charges''.
The solutions for this set of equations give the possible values
of the center positions. 
The asymptotic flatness condition implies
$ \sp{\R_\infty}{\I_\infty}=\sp{\ssigma\I_\infty}{\I_\infty}=1$.
%\label{eq689a}\end{align}

In practice, specific solutions are determined by giving a
particular, suitable, ansatz for the symplectic vector $\I$.
% as a function of the spacetime coordinates.
For  multicenter BPS solutions 
 we consider \cite{Bates:2003vx,Bellorin:2006xr,Fernandez-Melgarejo:2013ksa} a symplectic real vector $\I$ of the form
\begin{eqnarray}
\I&=&\I_\infty+ \sum_i \frac{q_i}{\vecr-\vecr_i}.
\end{eqnarray}
Thus $N=2$ solutions with $n_c$ centers are specified by 
$n_c+1$ symplectic vector $\I_\infty$, $q_i$ quantities. 

%Metric elements and physical parameters (ADM mass, horizon area, intercenter distances, angular momentum) are given in terms of this\I (or its ``components'' $\I_\infty$, $q_i$) in form of symplectic invariant relations.

The $n_v$ complex moduli $z_\infty^a$ 
and with them all other macroscopic parameters,
$M_{ADM},\Sigma^a$ and possibly some or all of $r_{ab}$,
%(and therefore the scalar charges) 
are complicated, implicit functions
of the  $2n_v+2$ real components of $\I_\infty$.
%...The  quantities $\I_\infty$ are not physical parameters.  
As we show in this work, the standard, ``non-physical'', $2n_v+2$ real components of $\I_\infty$ can be exchanged by ``contravariant'' components
which are directly physical quantities and which obeying 
the conditions of flat asymptoticity and integrability conditions above.
%...The quantities $\I_\infty,q_i$ are symplectic vectors. Physically equivalent solutions (with the same entropy, ADM mass,intercenter distances, etc) are generated by applying symplectictransformations on them. (...) It would be convenient to have neccesary and sufficient conditions expressed in terms of macroscopic and symplectic invariant quantities.
% dim:             2 nv+2,
% complex scalars: nv,
% real scalars:   2 nv 
% centers:        nc,
% real sa:            2 na=(2nv+2-2nc)
% masses, r_ij:       nc Mi,   nc(nc-1)/2 rij 
% parameters:    (Mi+rij+za+sa)= nc+nc(nc-1)/2+2nv+2 nv+2-2nc
% equations mi:     nc(nc-1)/2
% real equations za:      2nv  
%dof=parameters-equations: nc+2 nv+2-2nc=2nv+2-nc
%%
%%
Let us note that for a fixed configuration of $n_c$ centers, or charges,
 we have $n_c$ partial mass parameters $M_i$ 
and $n_c (n_c-1)/2$ intercenter parameters $r_{ij}$. 
In a model with 
$n_v$ complex scalars, the symplectic space is of dimension $2 n_v+2$...

{\bf 2. General Mass formulas.} Any real symplectic  vector 
$X$ of dim $2 n_v+2$ 
can be expanded in a basis of 
 charge vectors (or lineal combinations of them),  
and some additional $n_a$ (possibly zero) 
vectors, $s_a$ \cite{Fernandez-Melgarejo:2013ksa}.
% if needed to complete a basis,$n_a$ possibly equal to zero
In particular we can write,  in terms of 
eigenvectors of $\ssigmai$ 
\footnote{$w_i=(q_{nc},s_a)$,
the $s_a$ are chosen  such that $\sp{s_a}{\Pp q_i}=0$ for all the $q_i,s_a$. Projectors onto the 
$\pm i$ eigenspaces of $\ssigmai$ are 
 $\Ppm=(1\pm \ssigmai)/2$. $\Ppm^2=\Ppm$, $P_\pm P_\mp=0$. 
%Expansions of this type can be seen as alternatives to well known expansions $DZ...$.
},
\begin{eqnarray}
%X&=& \alpha^i \Pp q_i+\alpha^{\ibar} \Pm q_i+\lambda^a \Pp s_a+\lambda^{\bar a} \Pm s_a
X&=& \alpha^k \Pp w_k+\alpha^{\kbar} \Pm w_k
\label{eq03}
\end{eqnarray}
Associated to the basis we define the
 ``metric'' $g$ with components 
$g_{k\kbar}\equiv \sp{\Pp w_k}{\Pm w_k}$.
The submatrix
$g_{a{\bar b}} \equiv \sp{\Pp s_a}{\Pm w_b}$ 
can be choosen diagonal but
it is in general indefinite. 
The``contravariant'' components of $X$ with 
respect the metric $g$, are given by 
\begin{eqnarray}
a_j &=& \sp{X}{\Pp w_j}= g_{\ibar j}\alpha^\ibar,
%\quad a_a=a^a.
%,\\a_\jbar &=& \sp{X}{\Pm q_j}= g_{i\jbar}\alpha^i. 
\end{eqnarray}
In addition $ g_{i\jbar}g^{\jbar k}=\delta^{k}_{i}$. 
%$ g_{a\bbar}g^{\bbar c}=\delta^{c}_{a}$ ??.
Any  ``scalar'' product of the type $\sp{\Pp X}{\Pm X}$ 
is given by 
{%\small
\begin{eqnarray}
\sp{\Pp X}{\Pm X} &=& 
%\alpha_i \alpha^\ibar+\lambda_a\lambda^{\bar a}
\alpha_i \alpha^\ibar
=\alpha_i \alpha_\jbar  g^{i \jbar}
%=\alpha_i \alpha_\jbar  g^{i \jbar}+\lambda_a \lambda_{\bar b}  g^{a \bbar}.
\end{eqnarray}
}
The metric matrix $g_{i\jbar}$ can be decomposed in real (antisymmetric) and imaginary (simmetric) parts as follows:
$g_{i\jbar}\equiv (A+ i S)/2=(1/2) \sp{q_i}{q_j}+ (i/2) \sp{\ssigmai q_i}{q_j}.$
%\footnote{Explicit expressions for $X,Y$ are easily found.For example, for $A$  invertible,$X=2\left (A+S A^{-1}S \right )^{-1}$, $Y=A^{-1} S X.$.}}.

Let us apply the previous formulas to the vector $\I_\infty$.
%which is an essential part in the construction of our BH solutions.
The contravariant components  of it are given by 
well known physical parameters of the black-hole solution,
  $a_j(\I_\infty)=( N_j-i M_j)/2$ $j=1,n_c$ 
 and $2n_v+2-2 n_c$ additional complex parameters   
$a_a$ whose meaning will appear clear in what follows.

The asymptotic flatness condition becomes
$\sp{\ssigmai \Iinf}{\Iinf}=1=-2 i\sp{\Pp\Iinf}{\Pm\Iinf}$.
%
%Taking imaginary parts in both sides of this  product we arrive to
%\begin{eqnarray}1 &=& 2 \im{\alpha_i\alpha_\jbar}\re{ g^{i\jbar}}+2 \re{\alpha_i\alpha_\jbar}\im{ g^{i\jbar}}+ \lambda_a\lambda^\abar.\end{eqnarray}
%or , in terms of the contravariant components
%%\footnote{ 
%% $X^{ij}\equiv \re{ g^{i\jbar}}$,$Y^{ij}\equiv \im{ g^{i\jbar}}$}are respectively antisymmetric and symmetric matrices, $g^{i\jbar}\equiv X+ i Y$.
%The real matrices $X^{ij},Y^{ij}$ are given in the following way. 
%%{\small
%%\begin{eqnarray}1 &=& 2 \im{\alpha_i}\im{\alpha_j} Y^{ij}+4 \im{\alpha_j}\re{\alpha_i}X^{ji}\\&& +2 \re{\alpha_i}\re{\alpha_j}Y^{ij}+\lambda_a\lambda^\abar.\end{eqnarray}}
Writing this asymptotic condition in terms of contravariant components
and
taking into account that ( for $i=(1,n_c)$),  
$-2\im{\alpha_i}=M_i$, $2\re{\alpha_i}=N_i$
and defining $M_{ADM}=\sum_i M_i$, $m_i=M_i/M_{ADM}$, 
the asymptotic flatness condition becomes equivalent to the
mass relation
\begin{eqnarray}
1 &=& a M_{ADM}^2 +b M_{ADM}+c+\lambda_a\lambda^\abar
\label{eq01}
\end{eqnarray}
with\footnote{  $X^{ij}\equiv \re{ g^{i\jbar}}$,$Y^{ij}\equiv \im{ g^{i\jbar}}$
for $i,j=1,n_c$.}
$ a\equiv  (1/2) m_i m_j  Y^{ij}$,
$b= -  m_i N_j X^{ij}$,
$c = (1/2) N_i N_j Y^{ij}$.
The quantities $a,b,c$ depend on the symplectic 
products of charges (related to the BH angular momenta),
 the moduli at infinity (through 
the matrix $\ssigmai$ appearing in products 
$\sp{\ssigmai q_i}{q_j}$), the intercenter distances (appearing
in $N_i$, Eq.(\ref{eq5554})) and the relative parameters $m_i$. 
\footnote{The relation (\protect\ref{eq01}) is quadratic in 
$M$. The reality and 
positivity of $M$ imposes an upper bound on the contribution
of the extra vectors $s_i$
$ \lambda_a\lambda^\abar\le 1+(b^2-4ac)/4a$.
Eq.(\ref{eq01}) is also quadratic on the 
angular momentum components $J_{ij}\equiv \sp{q_i}{q_j}$
or in its module $J$. It can be written as 
$1= a M^2+b' J^2 M+ c' J^2 +\Delta^2$.
Similarly, reality of $J$  imposes 
conditions on the coefficients of the cuadratic expression.}
Eq.(\ref{eq01}) possibly admits solutions with 
$r_{ij}\to\infty$. In such a case $N_i\to 0 $ and we have 
($\Delta \equiv \lambda_a\lambda^{\bar a}$).
\begin{eqnarray}
1&=& a(m_i) M_\infty^2+ \Delta.
\end{eqnarray}
For fixed charges, the quantity $M_\infty^2$ 
depends on the configuration of relative masses $m_i$, it
is extremal (in fact a minimum)
for a configuration such that
($u\equiv (1,1,\dots)$
\begin{eqnarray}
m_{i,min} &=& \frac{Y^{-1}u}{u^t Y^{-1} u }
\simeq \frac{\sp{\ssigmai Q}{q_i}}{\sp{\ssigmai Q}{Q}}+O(J^2),
\end{eqnarray}
in this case $a_{max}(m_{i,min})=1/(u^t Y^{-1} u)$.

Eq.(\ref{eq01}) is complemented by additional consistency 
equations from the scalar fields. 
The values of the scalar fields at infinity are related 
to $\I_\infty$ by the expression 
$z^\alpha_\infty= Z^\alpha_\infty/Z^0_\infty$ where 
($i,j=(1,n_c)$) \cite{Fernandez-Melgarejo:2013ksa}
\begin{eqnarray}
Z_\infty&=& \Pm \I_\infty=
\frac{1}{2} \left (N_j-i M_j\right )g^{j\ibar}\Pm q_i+\lambda^a \Pm s_a.
\label{eq02}
\end{eqnarray}
This is an implicit equation including the quantities $z_\infty$ 
at both sides of the equation (as $\Pm=\Pm(z^\alpha_\infty)$).
Equation (\ref{eq02}) allows, if needed, 
to express the ``unphysical'' quantities $\lambda_a$ in 
terms of the scalar fields at infinity, masses, charges and 
intercenter distances. 
In the case that the $\lambda_a$ does not
appear, the equations (\ref{eq02})
 (one complex equation for 
each of the scalar fields) imposes additional constraints.

In the case the number of centers is large enough, one can 
select a subset of charge vectors to form a basis where 
to expand $\I_\infty$. This procedure can be repeated for 
different subsets generating different equations similar to
Eq.(\ref{eq01}) which should be simultaneously
satisfied.
%
%%%%%%%%%%%  XXXX ??????????????
%Other expansions, alternative to expression (\ref{eq03}) 
%are possible. Let us consider a basis formed 
%by $(\Ppm Q)$  and any needed number of additional auxiliar vectors 
%$s_a$, that is let us consider an expansion of the type
%$\I_\infty= \alpha^0 \Pp Q+\alpha^{\bar 0} \Pm Q+\lambda^a
% \Pp s_a+\lambda^{\bar a} \Pm s_a$. 
%The (only) contravariant coordinate is in this case 
%$a_0=(1/2)(N-i M_{ADM})$ where $N=0$ and $M$ is the total ADM mass.
%Imposing asymptotic flatness
% we arrive to an alternative quadratic mass relation
%\begin{eqnarray}
%1&=&  a M^2_{ADM}-\lvert \lambda\rvert^2,\quad a=1/\sp{\ssigmai Q}{Q}.
%\end{eqnarray}
%This relation can be written as 
%\begin{eqnarray}
%%M^2_{ADM}-\Sigma^2-W_{BH}&=&0.
%M^2_{ADM}&=&\sp{\ssigmai Q}{Q}\left(1+\lvert\lambda\rvert^2\right).
%\label{eq04}
%\end{eqnarray}
%%%Where
%%%$W_{bh}(q_i,z_\infty)=\sp{\ssigmai Q}{Q}$ and 
%%%$\Sigma^2=\Delta/a\equiv\Sigma^2(z_\infty^\alpha,\Sigma^\alpha)$.
%%
%%Thus, we have arrived to  a similar equation as  Eq.\ref{eqgibbons}.The dependence of $\Sigma^2$ in terms of scalar charges and infinity scalar fields is brought up by the eliminationof the parameters $\lambda^a$ from Eq.(\ref{eq04}) and a equation of the type (\ref{eq02}). 
%%??$Z_\infty= \Pm \I_\infty=-\frac{i}{2} M g^{j\ibar}\Pm Q+\lambda^a \Pm s_a$.
%%\label{eq02b}\end{eqnarray}

%%%%%%%%%%%%%%%%%%%%%%%%%%%%%%%%%%%

{\bf 3. Two center mass relations.} For the sake 
of concreteness, let us  consider now a particular, non-trivial case: 
a model with one scalar ($\bar n=2$) and two centers (with charge 
vectors $q_{1,2}$). 
The dimensionality of the sympletic space is $2 \bar n=4$, 
and the degrees of freedom 
$dof=2$. 
% according to Eq.(\ref{xxx}) 
%%are given by
%%\begin{align}A&=\sp{q_1}{q_2}\left(\begin{array}{cc}0 & 1\\ -1 & 0 \end{array}\right),&S &=\left(
%%\begin{array}{cc}
%% \sp{\ssigmai q_1}{ q_1} & \sp{\ssigmai q_2}{ q_1}\\
%% \sp{\ssigmai q_2}{ q_1} & \sp{\ssigmai q_2}{ q_2}
%%\end{array}
%%\right).
%%\end{align}
%%and the matrices $X,Y$ are 
%%\begin{eqnarray}X &=& \frac{-J}{2\det (g_{i\jbar})}
%%J\left(\begin{array}{cc}
%% 0 & 1\\-1 & 0 
%%\end{array}\right),\\
%%Y &=& \frac{-1}{2\det (g_{i\jbar})}
%%\left(
%%\begin{array}{cc}
%% \sp{\ssigmai q_2}{ q_2} & -\sp{\ssigmai q_2}{ q_1}\\
%% -\sp{\ssigmai q_2}{ q_1} & \sp{\ssigmai q_1}{ q_1}
%%\end{array}\right).
%%\end{eqnarray}
The hermitian matrix $ ( 2 i g_{i\jbar})$ is of signature $(1,1)$, 
therefore with negative determinant:
%$ \det ( 2 i g_{i\jbar})=\det( S)- \det(A)=det(S)- J^2<0$
$ \det ( 2 i g_{i\jbar})=det(S)- J^2<0$.
 $A,S$ are the real,imaginary parts 
of  $g_{i\jbar}$, in particular
$S_{ij}=\sp{\ssigmai q_i}{ q_j}$.
We have introduced the (signed) module, $J$, 
of the angular momentum
\footnote{We define $\vec J$ as
\begin{align}
\vec J=\sp{q_1}{q_2}\frac{\vecr_1-\vecr_2}{|\vecr_1-\vecr_2|}
\end{align}}.
The $\omega-$form compatibility equations 
or absence of NUT charge condition for this case read:
\begin{align}
 -N_2=N_1=\frac{\sp{q_1}{q_2}}{r}\equiv\frac{J}{r}.
\end{align}
We define in addition $Q=q_1+q_2$, $\atot=\sp{\ssigmai Q}{Q}$ 
%$a'=-det(S) (S^{-1})_{m_i m_j}$. 
and $M_0^2=1/ S^{-1}_{ij}{m_i m_j}$. 
The equation Eq.(\ref{eq01}) becomes for this two center case
\begin{eqnarray}
M^2 &=& M_0^2 \left (1+  \frac{J^2}{-\det(S)}\left(1+\frac{2M}{r}+\frac{A}{r^2}\right)\right).
\label{eq20}
\end{eqnarray}
The solution to this quadratic equation for $M$
has a  real and positive solution, assuming $\atot>0$,
only if $\det(S)<0$. 
%Previous equation implies $\det(S)<0$ or $0<\det(S)<J^2$. 
% In the next paragraphs we will see that the first option has to be choosen.....
In this case there exists a, unique,
  solution  for any intercenter distance $r$ and $J^2$. 
The quotient $J^2/(-\det(S))$ is  always positive
that implies that 
 $M^2>0$ if $M_0^2>0$.
%, or  $ (S^{-1})_{ij}m_i m_j>0$. This last inequality is not automatically guaranteed due to  the indefinitiness of  $S$.
%
The quantities $det (S),\atot, M_0^2$ 
depends on the charges and the moduli at infinity, in addition
$M_0^2$ depends on the relative masses of the centers.
%For quadratic prepotencials (for example axion-dilaton model or $CP^N$ models , $\ssigmai=\ssigma$, it dissapears the dependence with the moduli. 
%
In the limit of large intercenter distance, $r\to\infty$,
 the equation (\ref{eq20}) becomes
\begin{eqnarray}
M_\infty^2&=& M_0^2\left (1+\frac{J^2}{\lvert \det(S)\rvert}\right)
\end{eqnarray}
where $M_\infty$, if real,  is the total mass that would 
have such configuration of charges with an infinity intercenter distance.
At large distances $M\sim M_\infty+ J^2/(\lvert\det(S)\rvert r)$ while
at short distances the ADM mass behaviour is of the type 
$M\sim A/r+B r$.

%%
%% In the limits $J\to 0,J\to\infty$ the equation respectively becomes
%% \begin{eqnarray} M^2&\to &M_0^2, \quad J\to 0,\\ M^2&\to & -\frac{\atot}{2 r}-\frac{r}{2}+0(\frac{1}{J^2}). 
%%\end{eqnarray}

%%%%%%%%%%%%%%%%%%%%%
{\bf Smarr like relations.}
Let us use the mass differential $dM$ to define
different quantities  \emph{ à   la Smarr}
$$ dM\equiv\Omega d J+\Phi_i d q_i+ F dr,$$
in addition to an effective angular velocity $\Omega$ and 
electromagnetic potentials $\Phi_i$, the equation 
allows to define an effective ``force'', $F$,
 acting between the centers. 
%% dM/dr is a monotical function... with the 
%% same sign... whose...
This  force is always negative due to the 
 sign of $\det(S)$ enforced by Eq.(\ref{eq20}). 
At infinity $dM/dr \sim -J^2 M_0^2/(\lvert\det(S)\rvert r^2)\to 0$
%% ...Monotonically??, 
%%Tiene dM/dr un minimo??, probablemente no.
while at short distances $dM/dr\sim f_0+f_1/r^2$.

%%%%%%%%%%%%%%%%%%%%%%%%%%%%%%%%%
The positivity of $1/M_0^2=(S^{-1})_{ij}m_i m_j$ implies restrictions on the 
allowed values of the relative parameters $m_i$.
The allowed $ m_i$ are % ( $\atot>0,-\det(S)>0$ )
in an interval 
$m_{i,\text{min}}\pm \sqrt{\lvert\det(S)\rvert}/\atot$.
%%%%
The   relative masses that minimizes 
$M_\infty^2$ for fixed $J$ and $\ssigmai$ are given in this
case exactly by the expression
\begin{eqnarray}
m_{i,\text{min}}&=&\frac{\sp{\ssigmai q_i}{Q}}{\sp{\ssigmai Q}{Q}},
\end{eqnarray}
and the value of the total mass at minimum is given by
\begin{eqnarray}
(M_{0}^2)_{\text{min}} &=& \atot.
\end{eqnarray}
For this mass configuration we finally get ($\alpha =1/\lvert\det(S)\rvert$)
%%%\ref{eq20} 
\begin{eqnarray}
M^2 &=& \atot \left (1+  \alpha J^2\left(1+\frac{2M}{r}+\frac{A}{r^2}\right)\right).
\label{eq21}
\end{eqnarray}
%.....
Then 
%$(M_\infty{}^2)_{min}=\atot (\det(S)-J^2)/\det(S)$$=\atot \det (2 i g S^{-1})$.
$(M_\infty{}^2)_{min}=\atot \det (2 i g S^{-1})$.
At large distances 
$M\sim M_{\infty,min}+ J^2\atot/(\lvert\det(S)\rvert r)$ 
and   
$dM/dr \sim -J^2 \atot/(\lvert\det(S)\rvert r^2)\to 0$. 

As in the general case, 
these mass relations have to be
 complemented by the implicit equations for the 
moduli at infinity.

%%%%%%The expansion in terms of the total charge and one extra vector $s$ produces the mass relation\begin{eqnarray}M^2&=&...\end{eqnarray}(... relation of the extra parametersto the scalar fields and the scalar charges...).

%% For the strin-inspires, cubic case...

In conclusion we have derived 
new mass formulas for multicenter BPS 4D black holes.
They are obtained in the context of  
$N = 2,d = 4$ supergravity coupled to any number of vectors 
and with any number of BH centers.

%%...Non extremal mass formulas...... massless.....stringy 2 center, 3 center... 8 dimesions(...) the scalar fields at infinity and at the centers, interpolation formula, maybe it possible to  show that if the real part of the fields is positive in the three (Nc+1) points, then is positive in all the range, work in progress.. 

\emph{ Acknowledgements}
We acknowledge T. Ort\'{\i}n  and JJ. Fernandez-Melgarejo 
for many useful comments and suggestions.
This work has been supported in part by the Spanish Ministerio de 
Universidades and Fundacion Seneca (CARM Murcia) grants FIS2015-3454, PI2019-2356B and
the Universidad de Murcia project E024-018.

\newpage

\onecolumngrid

\appendix

\renewcommand{\baselinestretch}{1.1} 

\selectfont

\section*{SUPPLEMENTARY MATERIAL}

\subsection{ A particular example.}
\begin{figure}[ht]
\begin{tabular}{c}
\hspace{-0.97cm}\includegraphics[scale=0.65]{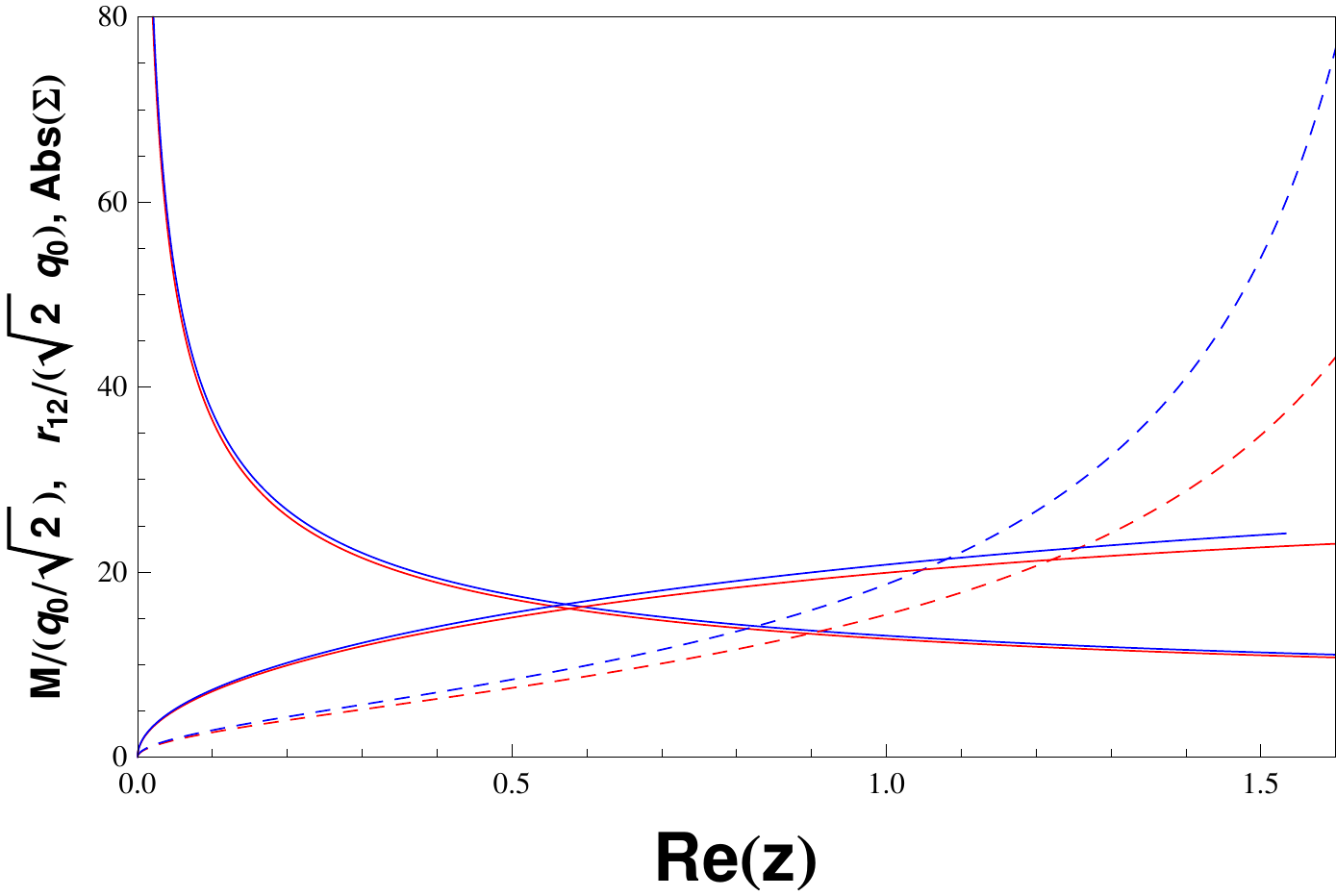}\\
\includegraphics[scale=0.65]{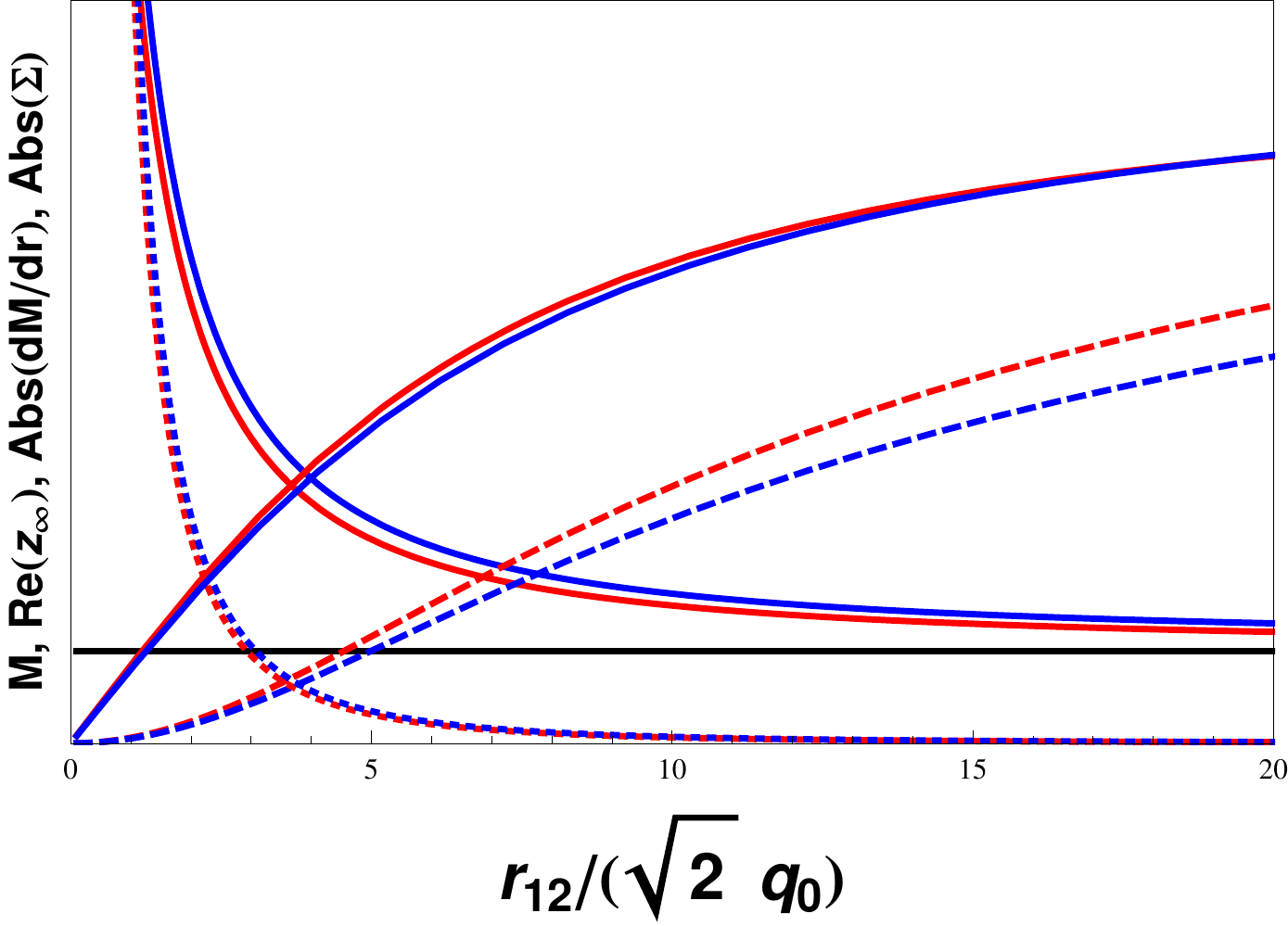}
\end{tabular}
\caption{
(A)(top). Dependence of $M_{ADM}$, $dM/dr,\re(z_\infty)$
 with respect to the intercenter distance 
$r$  for two configurations,
the minimal mass configuration and another $m_i=(9/19,10/19)$.
(B)(bottom)
$M_{ADM}$, $dM/dr$ and $r$ dependence
 with respect to the real part of the 
 scalar field at infinity, the dilaton, $\re{z_\infty}$.}
\label{aafg}
\end{figure}

Extremal mass relations (\ref{eq20}) or (\ref{eq21}) are valid 
for two center BHs in any N2 SUGRA.
In Fig.(\ref{aafg}) we present some explicit results for a 
 toy model with a complex scalar field $n = 2$ theory with prepotential
$F= -i X_0X_1$ (see further details in 
\cite{Bellorin:2006xr,Fernandez-Melgarejo:2013ksa}). 
For this quadratic prepotential the matrix 
$S$ is scalar independent. The only scalar of the theory  is 
$\chi+i e^{-\phi}\equiv -i z$.
The Kahler potential and scalar metric 
are ${\cal K}=-\log \re{z}, {\cal G}_{z\bar{z}}=(2 \re{z})^{-2}$. 
% This implies that \re(z)>0$.
%
Let us first take a configuration with
the charge sympletic vectors
$q_1=(1,8,0,-1 )q_0$ and
$q_2=(1,8,-4,1 )q_0$. 
In this case 
%%$\atot= 64 q_0^2$, $-\det(S)=272 q_0^4$, $J^2=144 q_0^4$.
$\atot>0, -\det(S)>0$.
The minimal ADM mass configuration  
corresponds to  relative masses
$m_i=M_i/M_{ADM}= (9/16, 7/16).$
The limiting mass is 
$M_\infty^2=\sp{\ssigma Q}{Q}\left(1-J^2/\det(S)\right)=8 \sqrt{26/17} q_0$.
%The range of allowed values ($M_\infty^2$) for $m_1$ is $...$.
The initial free parameters for this 
fixed configuration of charges are
  $M_1,M_2,r,z_\infty$ which have to 
satisfy three real equations 
(the mass relation, (\ref{eq20}) and 
a complex scalar equation of the type (\ref{eq02})).
If in addition we choose a minimal ADM mass
configuration it remains only one free parameter (see figure).
Let us take a second exemplary 
configuration with
$q_1=(1,8,0,-1 )q_0$ and
$q_2=(1,8,-8,0 )q_0$. In this case 
$\atot= 80 q_0^2$, $-\det(S)= 320 q_0^4$,
$J^2=0$.
Now the intercenter distance is unrestricted
while the scalar at infinity is fixed ($z_\infty=8$).
%The scalar charge is ...

\end{document}